# Streamlined Pin-Ridge-Filter Design for Single-energy Proton FLASH Planning

Chaoqiong Ma, Jun Zhou*, Chih-Wei Chang, Yinan Wang, Pretesh R. Patel, David S. Yu,
Sibo Tian and Xiaofeng Yang*

Department of Radiation Oncology and Winship Cancer Institute,

Emory University, Atlanta, GA, 30322, USA

*Email: xiaofeng.yang@emory.edu and jun.zhou@emory.edu

**Running title:** Streamlined Ridge Filter for FLASH-RT

**Manuscript Type:** Original Research

**Abstract**

**Background:** FLASH radiotherapy (FLASH-RT) with ultra-high dose rate has yielded promising results in reducing normal tissue toxicity while maintaining tumor control. Planning with single-energy proton beams modulated by ridge filters (RFs) has been demonstrated feasible for FLASH-RT.

**Purpose:** This study explored the feasibility of a streamlined pin-shaped RF (pin-RF) design, characterized by coarse resolution and sparsely distributed ridge pins, for single-energy proton FLASH planning.

**Methods:** An inverse planning framework integrated within a treatment planning system was established to design streamlined pin RFs for single-energy FLASH planning. The framework involves generating a multi-energy proton beam plan using intensity-modulated proton therapy (IMPT) planning based on downstream energy modulation strategy (IMPT-DS), followed by a nested pencil-beam-direction-based (PBD-based) spot reduction process to iteratively reduce the total number of PBDs and energy layers along each PBD for the IMPT-DS plan. The IMPT-DS plan is then translated into the pin-RFs and the single-energy beam configurations for IMPT planning with pin-RFs (IMPT-RF). This framework was validated on three lung cases, quantifying the FLASH dose of the IMPT-RF plan using the FLASH effectiveness model. The FLASH dose was then compared to the reference dose of a conventional IMPT plan to measure the clinical benefit of the FLASH planning technique.

**Results:** The IMPT-RF plans closely matched the corresponding IMPT-DS plans in high dose conformity (conformity index of < 1.15), with minimal changes in $V_{7Gy}$ and $V_{7.4Gy}$ for the lung (< 5%) and small increases in maximum doses ($D_{max}$) for other normal structures (< 3.2 Gy). Comparing the FLASH doses to the doses of corresponding IMPT-RF plans, drastic reductions of up to nearly 33% were observed in $D_{max}$ for the normal structures situated in the high-to-moderate-dose regions, while negligible changes were found in $D_{max}$ for normal structures in low-dose regions. Positive clinical benefits were seen in comparing the FLASH doses to the reference doses, with notable reductions of 18.4-33.0% in $D_{max}$ for healthy tissues in the high-dose regions. However, in the moderate-to-low-dose regions, only marginal positive or even negative clinical benefit for normal tissues were observed, such as increased lung $V_{7Gy}$ and $V_{7.4Gy}$ (16.4-38.9%).

**Conclusions:** A streamlined pin-RF design was developed and its effectiveness for single-energy proton FLASH planning was validated, revealing positive clinical benefits for the normal tissues in the high dose regions. The coarsened design of the pin-RF demonstrates potential advantages, including cost efficiency and ease of adjustability, making it a promising option for efficient production.

1. **Introduction**

FLASH radiotherapy (FLASH-RT), characterized by ultrahigh-dose-rate irradiation (typically ≥ 40 Gy/s), has shown promising outcomes for mitigating normal tissue toxicity while maintaining tumor control compared to conventional-dose-rate irradiation.[1-4] Small animal studies utilizing proton, electron or photon beams at FLASH dose rates have demonstrated superior sparing effects for various organs-at-risk (OARs), including the lung, brain, skin and abdominal tissues.[1, 2, 5-7] The beneficial effects induced with FLASH was also observed in the first clinical trial for a patient with T-cell cutaneous lymphoma, where irradiating with electron beams achieved favorable results outcome for both normal skin and the tumor.[8] Moreover, the FAST-01 trial provided evidence of the effectiveness of proton FLASH-RT for the human patients with multiple bone metastases.[9, 10] Besides, the clinical trials of FLASH-RT on examining protons for bone metastases in the chest[11] and electrons for skin melanoma metastases[12] are ongoing. Although the radiobiological mechanisms underlying the FLASH effect are not entirely understood, hypotheses suggest oxygen depletion and reactive oxygen species production may play a role.[13, 14] Recent studies also suggest a minimum dose threshold of 5-10 Gy, in addition to the dose rate threshold, for triggering the FLASH effect.[1, 7, 15]

Proton beams are preferred for FLASH-RT due to their precise tumor targeting and superior normal tissue sparing capabilities attributed to the narrow Bragg peaks (BPs).[13] In addition, the wide treatment range of the proton beams is another advantage for FLASH radiotherapy. Pencil beam scanning (PBS) is widely implemented for intensity-modulated proton therapy (IMPT) offering excellent dose conformity.[16] To obtain spread-out BPs (SOBPs) for PBS-IMPT, the proton beam energies are typically modulated by an energy degrader and energy selection system in a clinical cyclotron. This process the proton beam energies are typically modulated by an energy degrader and energy selection system in a clinical cyclotron. This process, however, significantly reduces the beam current and hence prevents the delivery of proton beams at FLASH dose rates.[17] To tackle this issue, a potential solution is to utilize the highest energy beam for FLASH planning, bypassing the need for the inefficient energy modulation system. Efforts have been made to employ high-energy proton transmission beams (TBs) to cover the target with the entrance dose region while delivering BPs outside the patient, demonstrating feasibility in attaining the FLASH effect.[18-21] Nonetheless, the main disadvantage of this approach is the potential overexposure of the normal tissues situated adjacent to the distal edge of the target to the high exit dose of the TBs.

Attempts have been taken to incorporate BPs in FLASH planning, given the aforementioned challenge associated with TBs. Lin *et al.* proposed a method that combines BPs of conventional dose rates and TBs for FLASH planning.[22] In particular, BPs were utilized to ensure central target dose coverage and normal tissue sparing beyond the target, while TBs were used to achieve FLASH dose rate coverage at the target boundaries. Furthermore, Ma *et al.* demonstrated that combining TBs with SOBPs, produced by spreading out the single-energy BPs using general step-shaped bar ridge filters (RFs), can improve OAR sparing while maintaining high target dose homogeneity for FLASH planning, in contrast to TB-only planning.[23]

In addition to improved OAR sparing compared to the TBs, BPs also offer enhanced relative biological effect due to their high linear energy transfer (LET) effect.[24] To maximize the potential of BPs for FLASH-RT, studies have explored the use of customized range compensators and uniform range shifters (RSs) to align the single-energy BPs to the distal edge of the target, indicating excellent OAR sparing with sufficient FLASH dose rate coverage.[25-27] Another avenue of exploration involves the implementation of pin-shaped RFs,[28, 29] also known as 3D range-modulators,[30] to produce a uniform single-energy SOBP that conforms to the target along each pencil beam direction (PBD). This approach has demonstrated feasibility for FLASH therapy accounting for dose and dose rate coverage as well as LET effect simultaneously.[28] Instead of generating uniform SOBPs within the target by the pin-RFs, Zhang *et al.* introduced a novel pin-RF designing method to obtain the pin-RFs capable of modulating the dose distribution along each PBD in the same manner as in PBS-IMPT, facilitating flexible manipulation of the dose distribution.[31] Specifically, a downstream energy modulation strategy[32] was adopted to generate a multi-energy IMPT plan with RS plates of various thicknesses downstream of the nozzle for energy modulation, which was subsequently translated into the pin-RFs. This method is referred to as inverse pin-RF designing method hereafter. Nevertheless, it is important to note that the aforementioned pin-RF designs encompass slender pins of widths only a few millimeters, and each pin can feature over10 steps, requiring a fine resolution of 0.1 mm.

As an alternative to the existing pin-RF designs, this study explored the feasibility of a streamlined pin-RF design of coarse resolution and sparsely distributed ridge pins for single-energy proton FLASH planning. Building upon the inverse pin-RF designing method, we proposed a PBD-based spot reduction process to generate a multi-energy IMPT plan based on the downstream energy modulation strategy (IMPT-DS), which can then be translated into the streamlined pin-RFs. Additionally, we commissioned a multi-energy beam model for downstream IMPT planning, allowing for the integration of proposed FLASH planning method within a treatment planning system (TPS).

## 2. Methods and materials

This study utilized a monoenergetic proton beam with an energy of 250 MeV, which is the energy used by a Varian ProBeam system in 'FLASH mode'. This mode can deliver the beam without requiring an energy degrader and energy selection system, achieving a beam current of up to 800 nA[33]. We developed an inverse planning framework integrated within a TPS, RayStation 10B (RaySearch Lab., Stockholm, Sweden), to design streamlined pin-RFs of coarse resolution and sparsely distributed ridge pins for proton

FLASH planning with single-energy beams. The aforementioned inverse pin-RF designing method[31] was employed in this work, which translates a multi-energy proton beam plan into the pin-RFs for a single-energy proton beam plan. The framework consists of two main parts:

(1) Generating a multi-energy proton IMPT plan based on the downstream energy modulation strategy (IMPT-DS). A nested PBD-based spot reduction process is conducted during the IMPT-DS planning to iteratively reduce the number of energy layers along each PBD and the total number of PBDs.
(2) IMPT planning with pin-RFs (IMPT-RF), which translates the IMPT-DS plan into streamlined pin-RFs (Figure 1(a)) and the single-energy beam configurations for delivering the plan in FLASH mode.

In the following sections, we firstly review the mechanism of the pin-RF for energy and intensity modulation for single-energy proton beams, followed by detailed descriptions of each component within the framework.

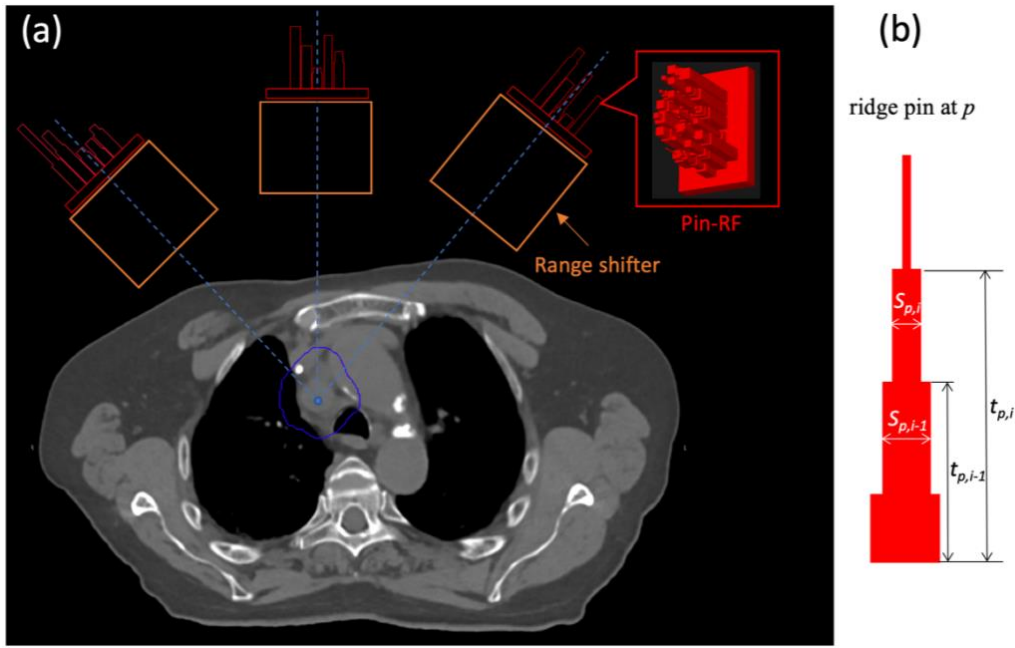

**Figure 1**. (a) A three-field arrangement for a lung treatment planning using pin-RFs and uniform RSs, with a 3D view of one pin-RF. The target is contoured in blue. (b) the cross-sectional shape of a ridge pin at a specific PBD, $p$.

### 2.1 Mechanism of the pin-RF

A pin-RF is built of multiple pyramid-shaped ridge pins mounting to a base (see Figure 1(a)). The mechanism of the pyramid-shaped ridge pin is to modulate the ranges (energies), referred to as water equivalent depths of the BPs, and fluences of an impinging monoenergetic proton pencil beam by using steps of varying thicknesses and surface areas.[34] As a result, the depth dose curve along this PBD is the weighted combination of the Bragg curves of varying BP depths. Specifically, as depicted in Figure 1 (b), the water equivalent thickness (WET), $t_{p,i}$, of step $i$ for a ridge pin at a specific PBD, $p$, corresponds to the Bragg curve $i$ with $i^{th}$ largest range (80% dose falloff), $R_{p,i}$, which can be determined by:

$$R_{p,i} = R_0 - t_{p,i}, \tag{1}$$

where $R_0$ is the range of the impinging monoenergetic beam. Let $s_{p,i}$ denote the width of the square surface of step $i$ at $p$. The relative weight, $\bar{w}_{p,i}$, of Bragg curve $i$, representing the proportion of the monoenergetic protons shooting through step $i$, can be recursively derived by equation (2) given the width, $s_{p,1}$, of the bottom step, referred to as the pin width.

$$\bar{w}_{p,i-1} = \frac{s_{p,i-1}^2 - s_{p,i}^2}{s_{p,1}^2} \tag{2}$$

$$\bar{w}_{p,i} = w_{p,i} / \sum_{i=1}^{N_E} w_{p,i} \tag{3}$$

According to the inverse pin-RF designing method adopted in this study, a ridge pin for a monoenergetic proton pencil beam can be translated from the energies and weights of the spots at a specific PBD of a multi-energy beam plan using equation (1) - (3) to obtain an equivalent weighted combination of the Bragg curves. Here $w_{p,i}$ and $N_E$ denote the weight of spot $i$ at $p$ and the number of spots/energies per PBD, respectively.

## 2.2 IMPT-DS planning

In this work, the multi-energy beam plan as a reference for the pin-RF design was acquired using the proposed IMPT-DS planning method. This planning method utilizes RS plates of various thicknesses placed downstream of the nozzle to modulate the energy of a monoenergetic proton beam,[31, 35] unlike the conventional IMPT (IMPT-CONV) planning method which uses an energy degrader and energy selection system. Similar to the IMPT-CONV planning, the energies and weights of the spots in IMPT-DS planning are obtained by solving a constrained optimization problem, achieving desired dose to the target while minimizing the dose delivered to the concerned normal structures. Let $N_P$ denote the total number of PBDs. The optimization problem can be formulated in (4). Here $\rho_o$ and $\rho_T$ are the penalties, $\bar{D}^o$ and $\bar{D}^T$ are the reference doses, for an OAR and a target, respectively. The dose $D_j$ at voxel $j$ is the weighted summation of the dose components, $d_j$, from $N_E$ spots at each PBD.

$$\min_{\mathbf{w}} f = \sum_{O \in OARs} \rho_O \sum_{j \in O} \{\max(0, D_j - \bar{D}_j^O)\}^2 + \sum_{T \in Targets} \rho_T \sum_{j \in T} \{\max(0, \bar{D}_j^T - D_j)\}^2$$

$$s.t. \quad D_j = \sum_{p=1}^{N_P} \sum_{i=1}^{N_E} w_{p,i} d_{j,p,i} \quad (4)$$

$$w_{p,i} \geq 0$$

It is worth noting that the use of RS plates can retain the enlarged beam divergence arising from multiple Coulomb scattering (MCS) of protons in the pin-RF, thereby ensuring that the energies and the corresponding weights from an IMPT-DS plan can be replicated by modulating a single-energy beam using the translated pin-RF.

### 2.2.1 Beam model commissioning

To facilitate IMPT-DS planning in RayStation, a single Gaussian beam source model was commissioned. This beam model consists of 11 single spot beams, which were generated by modulating the 250 MeV beam with RS plates of WETs decreasing from 10 to 0 cm with an interval of 1 cm. The material used for the RS plates was PMMA of density 1.19 g/cm$^3$. Therefore, this beam model can be implemented for treatment planning in a range of up to 10 cm. Specifically, a single-energy beam source model was firstly produced using TOPAS Monte Carlo (MC) toolkit[36] based on the measured beam data including the integrated depth dose (IDD) curve, in-air lateral spot profiles and absolute dose of the 250 MeV beam delivered by the ProBeam system at our facility.[37] Using the produced single-energy beam model, the IDD curves, the beam data for the other 10 energies were subsequently calculated by shooting the 250 MeV beams through the RS plates of aforementioned WETs. Finally, the automatic beam commissioning procedure in RayStation was conducted by taking the measured and simulated beam data for 250 MeV and energies < 250 MeV, respectively, as input to produce a multi-energy beam model for IMPT-DS planning. Excellent agreements between the input beam data and the simulated beam data using the GPU-based MC dose engine in RayStation were observed (see Figure S1 and S2 in the Supplementary Material).

It is important to mention that the distances from the bottoms of these RS plates to the isocenter remained constant, aligning with the distance from the base of the translated pin-RF to the isocenter (RF-to-isocenter distance) in IMPT-RF planning. This approach aims to preserve a constant inverse square effect, keeping consistency in the spot size after modulations by the RS plates and the pin-RFs. At the snout-to-isocenter distance of 42 cm for the ProBeam system, the RF-to-isocenter distance was chosen as 31 cm in this study ensuring a sufficient room for adding an additional RS between the pin-RF and patient surface to adjust the treatment range while avoiding collision between the pin-RF and the snout.

### 2.2.2 Nested PBD-based iterative spot reduction

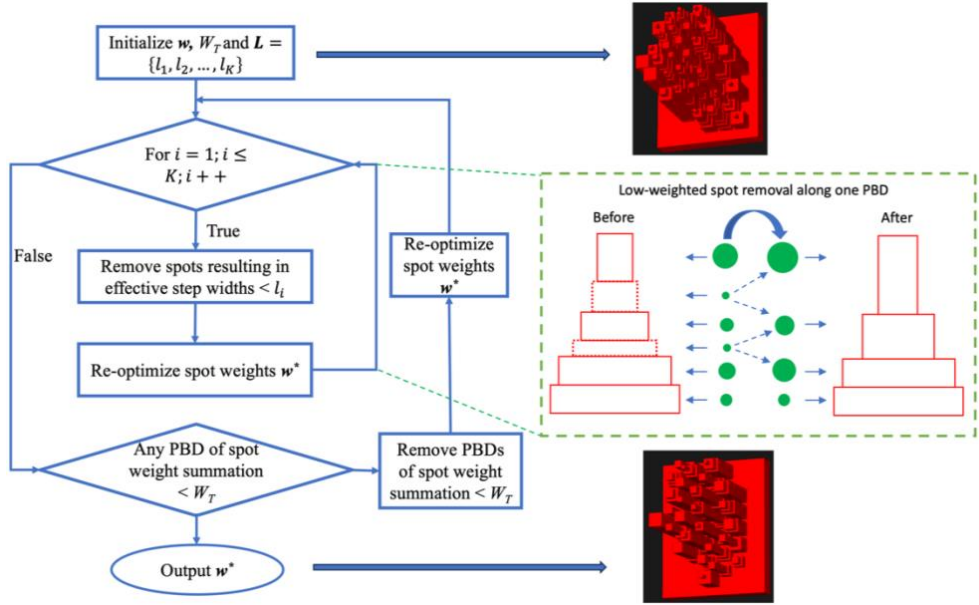

Figure 2. scheme of the nested PBD-based iterative spot reduction process. The green dashed box highlights an example of low-weighted spot removal for one PBD within one iteration of the inner loop. The green dots represent spots, with their sizes proportional to the spot weights. The resulting ridge pins before and after the spot removal are outlined in red. The corresponding steps of the ridge pin to the removed spots are contoured in dashed lines. Examples of the resulting pin-RFs for a field before and after the nested PBD-based iterative spot reduction process are also displayed.

The primary objective of the IMPT-DS planning is to create a plan with minimum number of PBDs and spots per PBD, resulting in streamlined pin-RFs with coarse resolution and sparsely distributed ridge pins for the IMPT-RF planning. Therefore, a nested PBD-based iterative spot reduction (ISR) process (Figure 2) was established for IMPT-DS planning, with an outer loop for removing PBDs with low spot weight summations, known as PBD reduction, and an inner loop for eliminating the low-weighted spots/energies for each PBD separately. In the inner loop, the low-weighted spots are iteratively eliminated, and their weights are redistributed to the adjacent spots at each PBD, which serves to increase the width differences between adjacent steps for the translated ridge pins, thereby coarsening the translated pin-RFs. Thus, we used the width difference between two adjacent steps, referred to as the effective step width, to determine the spots to be removed in the inner loop. For instance, the effective step width, $\bar{s}_{p,i}$, of step $i$ at $p$ can be calculated by equation (5). Larger $\bar{s}$ corresponds to higher $w$.

$$\bar{s}_{p,i} = s_{p,i+1} - s_{p,i} \qquad (5)$$

As is illustrated in Figure 2, an initial IMPT-DS plan with a set of optimal spot weights, $w$, is firstly generated by solving the optimization problem in (4). Note that a uniform RS is placed near the patient's surface for each field before initializing the plan aiming to pull the BPs of the available energies within the planning range. The WET, $T_{RS}$, of the RS, can be derived by:

$$T_{RS} = R_M - R_D - T_B \tag{6}$$

where $R_M, R_D, T_B$ are the range of 250 MeV beam, maximum distal depth of the target and the WET of the translated pin-RF base, respectively. Additionally, a threshold, $W_T$, of the spot weight summation per PBD and a set of effective step width limits, $\boldsymbol{L} = \{l_1, l_2, ..., l_K\}$, in ascending order ($l_1 < l_2 < \cdots < l_K$) are initialized. At each iteration of the outer loop, the low-weighted spots are iteratively removed by going through $\boldsymbol{L}$ in the inner loop and the PBDs of total spot weight $< W_T$ are subsequently eliminated. At each iteration of the effective step width limit $l$ of the inner loop, the spots of weights corresponding to the effective step widths $< l$ at each PBD are removed. Followed by the re-optimization of the weights for the rest spots. This process of spot removal and weight re-optimization is repeated to automatically distributed weights of the removed spots to the adjacent residual spots at each PBD until no spots need to be removed at $l$. Ultimately, any of the output spot weights $\boldsymbol{w}^*$ corresponds to an effective step width $\geq l_K$, and the spot weight summation at any PBD is not less than $W_T$.

**2.3 IMPT-RF planning with streamlined pin-RFs**

For single-energy IMPT-RF planning, the pin-RFs are derived from the spot weights of an IMPT-DS plan using equations (1) - (3). The center of each ridge pin aligns with the projected center of the corresponding PBD from the IMPT-DS plan at the pin-RF base, with a $T_B$ of 5 mm. As the parameters used in IMPT-DS planning, including spot spacing, effective step width limits ($\boldsymbol{L}$), and minimum spot weight summation per PBD ($W_T$) are linked to the pin-RF design and IMPT-RF planning, these parameters were chosen to meet the demands of the pin-RF design and IMPT-RF planning. In this study, the pin width $S_{p,1}$ was chosen as 8 mm and the pin spacing was set to 10 mm for the pin-RF design. Therefore, the projected pin spacing of 1.16 cm at the isocenter was utilized as the spot spacing for the IMPT-DS planning. It is worth mentioning that these two parameters for the pin-RF design were empirically chosen for our current beamline configuration, e.g., beam characteristics and the relative position of the pin-RF to the source and the isocenter, to assure dose conformity using the translated pin-RF for energy modulation in IMPT-RF planning. They may vary for different beamline configurations. Additionally, we coarsened the resolution of the step width for the translated pin-RFs to 1 mm by rounding the derived step widths using equation (2) to integral numbers of mm. With this intent in IMPT-DS planning, a set of effective step width limits ranging from 0.1 to 1 mm with a spacing of 0.1 mm was assigned to $\boldsymbol{L}$. Besides, a relatively high $W_T$ of 300 monitor units (MUs) was adopted in IMPT-DS planning ensuring the pin sparsity for the pin-RFs. This high $W_T$ also guarantees a high minimum spot weight for the resulting IMPT-RF plan, enabling deliverability of the IMPT-RF plan at a high beam current to achieve the FLASH effect.

The single-energy IMPT-RF plan is initialized with spots placed at the PBDs of the IMPT-DS plan. Subsequently, spot weight optimization is conducted by solving the formulated problem in (4). This additional spot weight optimization compensates for the dose deviation caused by varying proton MCS resulting from the pin-RF for IMPT-RF planning and the RS plates for IMPT-RS planning, thereby preserving high plan quality. for the IMPT-RF planning. In addition to setting $W_T$ to 300 MU for IMPT-DS planning, the minimum spot weight is also pushed above 300 MU in IMPT-RF planning to ensure the plan deliverability at a beam current of 500 nA and a minimum spot duration time of 0.5 ms.[38]

## 2.4 Evaluation of the FLASH effect

The FLASH effect of the proposed IMPT-RF planning method was evaluated by the FLASH effective dose of the IMPT-RF plans, which was quantified using the FLASH effectiveness model.[35] This is phenomenological model based on experimentally observed phenomena, such as potential dose rate and dose thresholds, and FLASH persistence time, providing a quantitative assessment of the FLASH effect for PBS proton therapy. This model operates on the principle that if the total dose delivered to a voxel of the normal tissues exceeds a dose threshold while the average dose rate is above a dose rate threshold during a time window, the FLASH effect is triggered. The FLASH effect is considered active during the FLASH trigger window and a subsequent persistence time. The dose delivered with the FLASH effect is assumed to be biologically less "effective" and thus multiplied by an effective factor (<1, reduced toxicity).[39] Note that no dose reduction is applied inside the GTV since tumor cells are assumed not to be affected by FLASH. In this work, the dose and dose rate thresholds, persistence time and the effective factor were set to 5 Gy, 40 Gy/s, 200 ms and 0.67, respectively, as in the previous study.[35] The scanning pattern was considered, and a scanning speed of 10 mm/ms, assumed by the Varian research group,[21] was used in calculating the FLASH effective dose.

## 2.5 Patient study

The proposed framework for FLASH planning was validated on three lung cancer patients who underwent proton stereotactic body radiation therapy (SBRT) at our institution. The size of clinical tumor volume (CTV) ranges from 45 to 129 cc. In this study, the prescription dose was set to 34 Gy delivered in a single fraction in accordance with the RTOG 0915 protocol for lung SBRT. Three beam angles were empirically chosen for each case, considering target dose conformity and avoidance of the OARs. The dose constraints for the OARs were adopted from this protocol for FLASH planning. For the target, which was the planning target volume (PTV) considering range uncertainty, the objectives for coverage were set as 95% receiving 100% of the prescription dose, $D_{95\%} = 34$ Gy, and $D_{99\%} \geq 30.6$ Gy. Besides, the maximum dose, $D_{max}$, of the PTV was limited to 140% of the prescription dose ($D_{max} \leq 47.6$ Gy).

The quality of the IMPT-RF plan was evaluated by comparing the IMPT-RF plan to the corresponding IMPT-DS plan in terms of the concerned dose indicators and the target dose conformity. The target dose conformity was quantified using the conformity index (CI), defined as the ratio of the total volume receiving the prescription isodose to the PTV volume.

The FLASH effect of each IMPT-RF plan was assessed using the FLASH effectiveness model, resulting in the FLASH dose distribution, referred to as RF-FLASH. The clinical benefit of FLASH achieved by the proposed planning method was measured by comparing the RF-FLASH to the reference dose distribution obtained from a conventional IMPT (IMPT-CONV) plan in terms of the concerned dose indicators as well as the mean dose of the GTV to PTV margin (PTV-GTV) for each case. This margin volume, potentially consisting of both healthy tissue and tumor cells and exposed to high doses of radiation, is especially intriguing as it may exhibit a pronounced and clinically relevant differential response if the FLASH effect operates at the cellular level.[35] To ensure consistency with the IMPT-DS planning, the same beam arrangement, prescription dose, target objectives, and constraints for the OARs were adopted for IMPT-CONV planning, following the guidelines outlined in the RTOG 0915 protocol.

Table 1. Absolute reductions in the dose indicators from the IMPT-RF to IMPT-DS plans (IMPT-RF vs. IMPT-DS), from the RF-FLASH to IMPT-RF plans (RF-FLASH vs. IMPT-RF) and from the RF-FLASH to IMPT-CONV plans ((RF-FLASH vs. IMPT-CONV) for the three cases

| | | Esophagus | | Great Vessels | | Heart | | Lung | | | Rib | | Skin | | Spinal Cord | | | Trachea | | PTV-GTV |
|---|---|---|---|---|---|---|---|---|---|---|---|---|---|---|---|---|---|---|---|---|
| | | $D_{5cc}$ (Gy) | $D_{max}$ (Gy) | $D_{10cc}$ (Gy) | $D_{max}$ (Gy) | $D_{15cc}$ (Gy) | $D_{max}$ (Gy) | $V_{7Gy}$ (cc) | $V_{7.4Gy}$ (cc) | $V_{20Gy}$ (%) | $D_{1cc}$ (Gy) | $D_{max}$ (Gy) | $D_{10cc}$ (Gy) | $D_{max}$ (Gy) | $D_{0.35cc}$ (Gy) | $D_{1.2cc}$ (Gy) | $D_{max}$ (Gy) | $D_{4cc}$ (Gy) | $D_{max}$ (Gy) | $D_{mean}$ (Gy) |
| Case 1 | IMPT-RF vs. IMPT-DS | -2.5 | 0.5 | -1.1 | -3.2 | -0.2 | 0.1 | -8.8 | -7.0 | -0.3 | -1.6 | -2.5 | 0.0 | -2.4 | -1.6 | -2.2 | -2.5 | 1.0 | 0.0 | |
| | RF-FLASH vs. IMPT-RF | 3.6 | 11.2 | 10.6 | 12.6 | 0.0 | 3.1 | 35.9 | 35.6 | 1.9 | 8.4 | 10.1 | 2.8 | 5.7 | 0.0 | 0.0 | 0.0 | 11.1 | 11.1 | |
| | RF-FLASH vs. IMPT-CONV | -3.9 | 11.4 | 8.6 | 9.6 | -1.9 | -5.4 | -48.8 | -44.6 | 0.7 | 1.2 | 5.3 | 0.6 | 0.4 | -2.0 | -2.7 | -3.5 | 11.7 | 10.5 | 8.6 |
| Case 2 | IMPT-RF vs. IMPT-DS | -0.1 | -0.2 | -0.7 | -0.8 | 0.3 | 2.1 | -10.3 | -10.4 | -0.5 | -2.1 | -2.3 | -0.5 | -1.7 | 0.4 | 0.5 | 0.1 | -0.2 | 0.2 | |
| | RF-FLASH vs. IMPT-RF | 0.0 | 0.0 | 2.9 | 7.7 | 0.4 | 3.5 | 28.4 | 30.5 | 4.5 | 11.8 | 12.2 | 3.7 | 4.9 | 0.0 | 0.0 | 0.0 | 0.0 | 0.0 | |
| | RF-FLASH vs. IMPT-CONV | -0.2 | -0.5 | -5.0 | 7.0 | -5.8 | 0.6 | -37.9 | -34.2 | 2.8 | 10.4 | 9.7 | -0.5 | 1.0 | -0.3 | -0.6 | -1.2 | -0.5 | -1.2 | 10.6 |
| Case 3 | IMPT-RF vs. IMPT-DS | -1.6 | -1.6 | -0.7 | 2.1 | -2.0 | -0.1 | -8.7 | -8.6 | -0.2 | -0.8 | -0.7 | -0.1 | -0.7 | -1.5 | -1.4 | -1.2 | 0.0 | -0.5 | |
| | RF-FLASH vs. IMPT-RF | 0.0 | 0.7 | 3.1 | 9.0 | 1.1 | 3.6 | 19.2 | 21.0 | 4.9 | 7.0 | 8.6 | 3.0 | 4.3 | 0.0 | 0.0 | 0.0 | 0.0 | 0.0 | |
| | RF-FLASH vs. IMPT-CONV | -2.3 | -6.7 | -5.0 | 10.9 | -5.0 | -2.9 | -46.2 | -37.7 | 3.6 | 4.5 | 7.6 | 0.5 | 2.3 | -3.4 | -4.4 | -5.0 | 0.0 | -0.6 | 10.7 |

## 3. Results

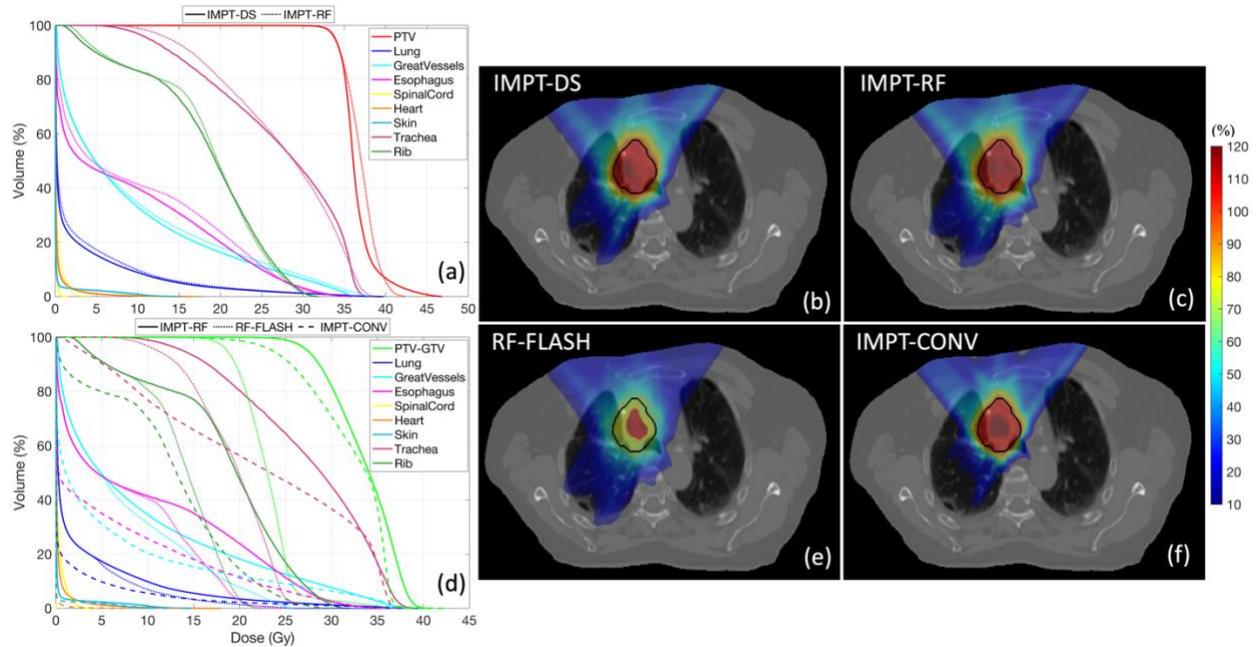

Figure 3. Dosimetric results for Case 1. (a) DVH comparisons between the IMPT-DS and IMPT-RF. 2D dose distributions of (b) IMPT-DS and (c) IMPT-RF. (d) DVH comparisons between IMPT-RF, RF-FLASH and IMPT-CONV. 2D dose distributions of (e) RF-FLASH and (f) IMPT-CONV. The PTV is contoured in black. The dose scale in isodose plot ranges from 10% to 120% of the prescription dose.

Figure 3 (a) illustrates the dose volume histograms (DVHs) of the IMPT-DS and IMPT-RF plans for Case 1. The 2D dose distributions of the corresponding plans for the same case can be seen in Figure 3 (b) and (c). For the other two cases, the DVHs and the 2D dose distributions of the two plans can be found in Figure S3 and S4 in the Supplementary Material. Both plans demonstrate high dose conformity to the target, with the 95% isodose surface closely aligning with the PTV. However, the IMPT-RF plan shows larger low-dose spill in the region adjacent to the distal edge of the target compared to the IMPT-DS plan. This could be attributed to the enlarged spot size/penumbra resulting from the MCS of the pin-RFs used in the IMPT-RF plan. For each case, both plans achieved high CIs of $\leq 1.15$, with a small CI difference $\leq 0.03$, indicating similar target dose conformity between the two plans. To further quantify of the plan quality, the relevant dose indicators of the concerned OARs according to the RTOG 0915 protocol were calculated for both plans of each case. The resulting reductions in these dose indicators from the IMPT-RF to IMPT-DS plans are summarized in Table 1, while a detailed breakdown of the dose indicators can be found in Table S1 in the Supplementary Material. For the OARs other than the lung, the relevant dose indicators are maximum doses for a point or small volume within the OARs, which increased (negative numbers represent increases in those parameters) by no more than 3.2 Gy in the IMPT-RF plans compared to the IMPT-DS plans. For the lung, $V_{7Gy}$ and $V_{7.4Gy}$ were slightly elevated by < 11 cc (5%) in the IMPT-RF plans compared to the IMPT-DS plans. The difference in $V_{20Gy}$ for the lungs was negligible between the two plans. Therefore, the IMPT-DS plans can be highly replicated by the corresponding single-energy IMPT-RF plans.

Figure 3 (d) an (e) display the DVHs and the 2D dose distribution of the RF-FLASH corresponding to the IMPT-RF plan for Case 1, respectively. The DVHs and 2D dose distributions of the RF-FLASH for the other two cases are displayed in Figure S3 and S4. Compared to the dose distribution of the IMPT-RF plan, referred to as original dose, a significant dose reduction was observed in the region surrounding the GTV for the RF-FLASH (FLASH dose). Since no FLASH effect was applied to GTV, as previously mentioned, the GTV dose of the FLASH dose was identical to that of the original dose. The relevant dose indicators of the OARs for the FLASH doses were extracted and are presented in Table S1. The resulting reductions in these dose indicators induced by FLASH are summarize in Table 1. For all three cases, dramatic reductions of $\geq 30.6\%$, $\geq 29.6\%$, ranging from 18.7-22.8% and 22.1-32.4% in $D_{max}$ for great vessels, rib, heart and skin, respectively, were achieved with the FLASH doses compared to the original doses. Additionally, the $D_{max}$ for esophagus and trachea drastically dropped by nearly 33% for Case 1, which is the maximum FLASH effect that can be achieved at the effective factor of 0.67, when comparing the FLASH dose to the original dose. These normal structures, which underwent a substantial FLASH effect, were situated near or overlapping with the PTVs and subjected to high (roughly above two thirds of the prescription dose) to moderate (between one third and two thirds of the prescription dose) original doses. In contrast, with the FLASH effect, the lung exhibited moderate decreases of 9.1-17.8% in $V_{7Gy}$ and $V_{7.4Gy}$, and a marginal drop of < 5% in $V_{20Gy}$. Similarly, negligible changes of within 0.7 Gy in $D_{max}$ were observed for the spinal cord of all cases, and for the esophagus and trachea of Case 2 and 3, which were subjected to low-dose (approximately below one third of the prescription dose) regions. The dependence of the magnitude of the FLASH effect on the dose regions can be clarified by examining the mechanism of FLASH triggering. As the FLASH triggering window is determined by the dose and dose rate thresholds, regions exposed to high to moderate physical doses tend to experience longer FLASH triggering widows. As a result, the FLASH effect provides greater benefits to the normal tissues in high-to-moderate-dose regions compared to those in low-dose regions.

To assess the clinical benefit of this FLASH planning method, a comparison was conducted between the FLASH dose and the reference dose of the IMPT-CONV plan. In Figure 3 (d), the DVHs of the IMPT-CONV plan for Case 1 are shown, while Figure 3 (f) displays the 2D dose distribution. Similar data for the other two cases can be found in Figure S3 and S4. The relevant dose indicators of the OARs and PTV-GTV of all cases for the IMPT-CONV plans are provided in Table S1, and the reductions observed in these dose indicators when comparing the FLASH and reference doses are listed in Table 1. A pronounced clinical benefit induced by FLASH effect was observed in the high-dose regions: dramatic reductions of $\geq 26.1\%$ and 18.4-27.9% in $D_{max}$ for great vessels and rib, respectively, across all cases, and 33.0% in $D_{max}$ for esophagus and trachea of Case 1 were achieved with the FLASH dose in contrast with the reference dose. Furthermore, the $D_{mean}$ for PTV-GTV drastically dropped by $\geq 26.7\%$ for all cases when comparing the FLASH dose to the reference dose. Thus, the proposed FLASH planning method can achieve substantial positive clinical benefit to the healthy tissues located in the high-dose regions. Note that the dose regions mentioned here were determined in accordance with the reference dose. Owning to the reduced spot sizes and enhanced degrees of freedom, the IMPT-CONV plan exhibited superior dose conformity compared to the IMPT-RF plan, leading to improved protection for normal tissues located in the moderate-to-low-dose regions. As a result, the FLASH dose only led to marginal reductions, with < 2.3 Gy in $D_{max}$ for the skin, and < 3.6% in $V_{20Gy}$ for the lung in all cases, when compared to the reference dose. However, negative clinical benefit for normal tissues in the moderate-to-low-dose regions can also be correlated with this FLASH planning technique. For instance, the FLASH dose was associated with an increase of 16.4-38.9% in $V_{7Gy}$ and $V_{7.4Gy}$ for the lung of all cases compared to the reference dose. Additionally, increments of up to 6.7 Gy in $D_{max}$ were induced for the esophagus and spinal cord of Case 3, as well as for the spinal cord and heart of Case 1 by the FLASH dose compared to the reference dose. Therefore, the proposed FLASH planning method may lead to only marginal positive clinical benefit or potentially negative clinical benefit for normal tissues in the moderate-to-low-dose regions.

## 4. Discussion

In this study, we established a novel inverse planning framework to design streamlined pin-RFs for proton FLASH planning using single-energy beams. Utilizing the inverse pin-RF designing method, we translated the pin-RFs for single-energy FLASH planning from a multi-energy proton beam plan, with the goal of obtaining the optimal weighted summation of the BPs, rather than a uniform SOBP[28, 29] in the target along each PBD, which allows for more flexible manipulation of the dose distribution. The nested PDB-based spot reduction scheme in the framework reduces spots per PBD and the total number of PBDs for the multi-energy proton beam plan, resulting in the streamlined pin-RFs with coarser resolution, fewer steps per ridge pin (no more than four steps per ridge pin), and fewer ridge pins in contrast with the pin-RF design of the representative pilot study.[31] Thus, the streamlined design can potentially simplify the manufacturing process (e.g., 3D printing) by reducing precision requirements and manufacturing time. Another major contribution of this study is the commissioning of a multi-energy beam model in RayStation for IMPT-DS planning. This allows for the integration of the proposed inverse planning framework within the TPS, facilitating efficient and user-friendly implementation in FLASH planning.

The single-energy IMPT-RF plan using the pin-RFs translated from the corresponding multi-energy IMPT-DS plan can replicate the IMPT-DS plan with high target dose conformity. There are three essential factors that contributes to this achievement. One of the key factors is maintaining a consistent RS-to-isocenter distance for energy modulation in the IMPT-DS planning and matching the RF-to-isocenter distances to that RS-to-isocenter distance for IMPT-RF planning. This approach ensures a consistent inverse square effect for the beams after being modulated by both the RS plates and the pin-RFs, resulting in consistent spot sizes and similar plan quality between the IMPT-DS and IMPT-RF plans. The second key factor is coordinating the appropriate pin width and pin spacing for the pin-RF design. In this work, we empirically chose the pin width of 8 mm and the pin spacing of 10 mm to assure independent and efficient beam modulation by the ridge pin along each PBD given our current beam characteristics, where the Gaussian-shaped spot profile for 250 MeV exhibited a standard deviation of 3.4 mm. The 10 mm pin spacing also assured that the corresponding spot spacing for IMPT-RF planning remained below 1.8 cm, thus preventing ripple effects in the dose distribution.[31] The third key factor involves implementing an additional beam intensity modulation for the IMPT-RF plan, which accounts for the deviation caused by the different MCS of protons resulting from the pin-RFs in IMPT-RF planning and the RS plates in IMPT-DS planning, thereby ensuring high dose conformity in the IMPT-RF plan.

The FLASH effect for the single-energy IMPT-RF plan were measured using the phenomenological FLASH effectiveness model. This model takes into account the dose and dose rate thresholds for determining FLASH triggering windows, and the FLASH persistence after triggering based on observations from previous studies to calculate the biologically effective dose in the presence of the FLASH effect. As the initial radiobiological studies suggest a dose rate threshold of $>40$ Gy/s for FLASH triggering, attempts have been made to quantify the dose rate for proton PBS plans using the dose-averaged dose rate (DADR) metric,[19] which calculates the voxel-wise instantaneous dose rate of each spot and takes the dose-weighted average of these dose rates. This metric may overestimate the dose-rate effect since it ignores dead times between spots. The average dose rate (ADR) metric[21] addresses this limitation by including dead times in the voxel-wise average dose rate. Nonetheless, simply using either dose rate metric for FLASH evaluation assumes that the dose in a voxel is either delivered fully with or without FLASH, which may not accurately represent the actual dose rate fluctuations during a proton PBS delivery. In addition to the dose rate threshold, the presence of a minimum dose threshold of 5-10 Gy[1, 7, 15] to trigger the FLASH effect and a FLASH persistence time of 200-500 ms[40] after triggering have been observed by recent studies. The FLASH effectiveness model takes all these parameters into consideration to extract the dose portions delivered in FLASH for each voxel and estimate the potentially reduced toxicity/dose for normal tissues during a PBS delivery, providing a more comprehensive reflection of the FLASH effect along with the PBS delivery dynamics of the proposed FLASH planning technique.

By estimating the FLASH dose of the IMPT-RF plan using the FLASH effectiveness model, we were able to evaluate the clinical benefits of the proposed single-energy planning method. This was done by directly comparing the FLASH dose to the reference dose obtained from the IMPT-CONV plan. Significant dose reductions of up to nearly 33% were achieved for the normal tissues situated in the high dose regions, leading to substantial positive clinical benefits. The accomplishment of such high FLASH dose effect can be attributed in part to the use of pin-RFs for single-energy PBS to deliver an entire modulated BP at each PBD, providing increased dose rates. Additionally, the employment of a hypofractionation scheme to deliver 34 Gy in one fraction helped push the dose above the dose threshold for FLASH triggering.[35] Another contributing factor was applying a relatively high beam current of 500 nA to increase the dose rate above the dose rate threshold for FLASH triggering. Noteworthy is that in IMPT-RF planning, the choice of the lowest achievable minimum spot duration time of 0.5 ms granted a minimum MU of 300, ensuring both high plan quality and plan deliverability. Despite the potential benefits of FLASH dose reduction induced by this single-energy PBS technique, it is important to acknowledge the negative impact on healthy tissues located in moderate-to-low dose regions due to the degraded dose conformity.

The limitation of this study should be noted. A relatively large RF-to-isocenter distance of 31 cm was used in this work to leave sufficient space for additional uniform RSs to pull back the BPs of the 250 MeV beams to the treatment range. For instance, the thicknesses of the uniform RSs simulated using PMMA (1.19 g/cm$^3$) were up to 23 cm (WET of ~27 cm) for the IMPT-RF plans. Such large RF-to-isocenter distance can (1) cause large spot size due to the amplified impact of MCS induced by the uniform RSs and pin-RFs at a high inverse square effect, thereby diminishing dose conformity; (2) limit the space (< 11 cm) for pin-RFs, thus limiting the treatment range for this technique. This can be a general issue for the single-energy planning techniques using BPs of high-energy beams for FLASH-RT. To address this issue, alternative materials of higher density and comparable MCS effect to PMMA for the uniform RSs need to be explored in order to reduce the gap between the pin-RF and the patient surface for improved dose conformity and thus better protection for OAR in the moderate-to-low dose regions.

This preliminary study aimed to explore the feasibility of the streamlined pin-RF design for single-energy FLASH planning. The use of coarsened pin step widths to the integer number of mm in this work paved the way for producing pin-RFs by assembling a few bricks of desired widths and thicknesses selected from a set of limited number of reusable bricks, rather than solely relying on 3D printing. The utilization of reusable bricks has the potential to accelerate the production process and reduce the cost of pin-RFs, in comparison to 3D printing. Furthermore, the assembled pin-RFs can be easily adjusted by replacing a few bricks as needed for plan adaptation during a treatment course. Henceforth, we will cover the design of the set of reusable bricks for pin-RF assembly, as well as the application of the assembled pin-RFs for proton adaptive therapy study. In addition, the dosimetric impacts of the uncertainties of the pin-RF and spot positions will be explored as part of our future work. We also plan to incorporate the FLASH effect, quantified by the FLASH effectiveness model, in the plan objectives to fully exploit the potential of the proposed pin-RF designing method for FLASH treatment.

## 5. Conclusions

In this study, we validated the effectiveness of the streamlined pin-RF design for single-energy proton FLASH planning. To evaluate the clinical benefits of this FLASH planning technique, we conducted quantitative comparisons between the FLASH doses of the IMPT-RF plans and the reference doses from the corresponding conventional IMPT plans. The results demonstrated positive clinical benefits of the proposed FLASH planning technique for the normal tissues in the high dose regions. The coarsened design for the streamlined pin-RF has the potential to enable efficient production at a reduced cost, as well as easy adjustability.

**Conflicts of interest**

The authors have no conflicts to disclose.

**Acknowledgment**

This research is supported in part by the National Cancer Institute of the National Institutes of Health under Award Number R01CA215718 and R01EB032680.